\documentstyle[12pt]{article}
\title{Effective Action in ${\cal N}=2,4$ Supersymmetric Yang-Mills Theories
}

\author{A.T. Banin\footnote{atb@math.nsc.ru}${}^{\;\;a}$, I.L.
Buchbinder\footnote{joseph@tspu.edu.ru}${}^{\;\;b}$ , N.G.
Pletnev\footnote{pletnev@math.nsc.ru}${}^{\;\;a}$}

\date{{\it ${}^{\;\;a)}$Institute of Mathematics, Novosibirsk, \\ 630090, Russia,\\
\vspace{0.7cm} ${}^{\;\;b)}$Department of Theoretical Physics\\
Tomsk State Pedagogical University\\ Tomsk 634041, Russia} }

\begin{document}
\begin{titlepage}
\maketitle
\begin{abstract}
We review the approach to calculation of one-loop effective action in
${\cal N}=2,4$ SYM theories. We compute the non-holomorphic
corrections to low-energy effective action (higher derivative
terms) in ${\cal N}=2$, $SU(2)$ SYM theory coupled to
hypermultiplets on a non-abelian background for $R_{\xi}$-gauge
fixing conditions. A general procedure for calculating the gauge
parameters depending contributions to one-loop superfield
effective action is developed. The one-loop non-holomorphic
effective potential is exactly found in terms of the Euler
dilogarithm function for a specific choice of gauge parameters.We
also discuss the calculations of hypermultiplet dependence of
${\cal N}=4$ SYM effective action.
\end{abstract}
\thispagestyle{empty}
\end{titlepage}

\newcommand{\be}{\begin{equation}}
\newcommand{\ee}{\end{equation}}
\newcommand{\bea}{\begin{eqnarray}}
\newcommand{\eea}{\end{eqnarray}}


\twocolumn
\section{Introduction}

It is well-known that low-energy effective action of ${\cal N}=2$
supersymmetric Yang-Mills theories is determined, in purely gauge
superfield sector, by two effective potentials. The leading
correction is given by holomorphic potential ${\cal F}({\cal W})$,
and the next-to-leading correction is written in terms of
non-holomorphic potential ${\cal H}({\cal W}, \bar {\cal W})$
where ${\cal W}$ and $\bar {\cal W}$ are ${\cal N}=2$ superfield
strengths (see e.g. the review [1]). ${\cal N}=2$ supersymmetry
strongly restricts the form of holomorphic potential what was
demonstrated by Seiberg and Witten for $SU(2)$ SYM model in the
Coulomb branch of inequivalent vacua in which the low energy
theory has unbroken $U(1)$ gauge factors [2]. An extension of this
result for various gauge groups and coupling to matter was given
in Ref. [3] (see also the review [4]). General form of holomorphic
potential for an arbitrary ${\cal N}=2$ model is now well
established.

Computation of the non-holomorphic potential is more delicate, and
a general form of ${\cal H}({\cal W}, \bar{\cal W})$ is still
unknown although some contributions to ${\cal H}$ were obtained
for special cases. In ${\cal N}=2$ superconformal invariant models
and the ${\cal N}=4$ SYM theory the non-holomorphic potential has
been found in the Coulomb phase [5] - [9]. Here all beta functions
vanish, and the evolutions under the renormalization group is
trivial. This effective potential is turned out to be an exact
solution of the ${\cal N}=4$ SYM theory, its explicit form is
given only by one-loop contribution, any higher-loop or instanton
corrections are absent [6], [9] - [11].
However
all above results correspond to Abelian background ${\cal W}$ and
$\bar{\cal W}$ for the theory, living on a point of general
position of the moduli space, where one has the symmetry-breaking
pattern: $SU(N)\rightarrow U(1)^{N-1}$ and all physical quantities
vary smoothly over the moduli spaces. As to a non-abelian
background, the non-holomorphic potential was found only for very
special cases in Refs. [5] - [12].

One of the basic approaches to evaluating the effective action is
the derivative expansion [15]. This approach allows one to get the
effective action in the form of a series in derivatives of its
functional arguments. Within ${\cal N}=1$ supersymmetric
derivative expansion, the leading contributions to the effective
action are formed by the so-called K\"{a}hlerian and chiral
superfield effective potentials [13], [14]. We point out that the
K\"{a}hlerian effective potential naturally arises in ${\cal N}=2$
SYM models if ones formulate these models in terms of ${\cal N}=1$
superfields [16] and, as a result, it allows to construct the
potentials ${\cal F}({\cal W})$ and ${\cal H}({\cal W}, \bar{\cal
W})$ on its base.

Another line of current study of the effective action in extended
SUSY theories is associated with a realization of these theories
on the world volume of branes. Such a realization provides a dual
description of low-energy field dynamics in terms of D-brane
theory. Webs of intersecting branes as a tool for studying the
gauge theories with reduced number of supersymmetries have been
introduced in Ref. [17]. The five-brane construction has been
successfully applied to a computation of holomorphic (or rather
BPS) quantities of the four-dimensional supersymmetric gauge
theory (see Refs. [18], [19]). The five-brane configurations
corresponding to these ${\cal N} = 1$ supersymmetric gauge
theories encode the information about the ${\cal N} = 1$ moduli
spaces of vacua. The non-holomorphic quantities such as higher
derivative terms in ${\cal N}=2$ theories and the K\"ahlerian
potential of ${\cal N}=1$ supersymmetric gauge theories are of
special interest since they are not protected by supersymmetry. It
was shown that the K\"ahlerian potential on the Coulomb branch of
${\cal N}=2$ theories is correctly reproduced from the classical
dynamics of M-theory five-brane. As to the non-holomorphic
contributions to low-energy effective action, such as the higher
derivative terms, a correspondence between string/brane approach
and four-dimensional ${\cal N}=2$ supersymmetric Yang-Mills
theories beyond two-derivative level, is not completely
established (see e.g. Refs. [18, 19, 20, 21]).

In this paper we discuss some aspects of structure of the
non-holomorphic effective potential for non-abelian background in
order to pay attention on a problem of its gauge dependence. This
fact is related to the parameterization non-invariance of the
conventional effective action (see e.g. Ref. [22]) and leads to a
number of different effective actions corresponding to one
classical action. But any gauge-fixing condition is equal to a
redefinition of fields in each order of the effective action loop
expansion [30]. The gauge dependence of an effective action for
Yang-Mills theories is well-known problem for many years [22,30].
But for  ${\cal N}=1,2,4$ super-YM theories such a problem has not
been considered in detail. A supersymmetric generalization of
$R_{\xi}$-gauge seems to be a good tool for studying
gauge-dependence in SYM theories. Such a generalization was first
suggested in [28]. We present an extended supersymmetrical
$R_{\xi}$-gauge for SYM models within background field method. The
choice of a gauge fixing term in spontaneous broken non-abelian
gauge theories is of basic technical importance. It is known that
the use of the $R_{\xi}$-gauge became a major step in the proof
that Yang-Mills models are unitary, on-shell gauge-independent and
renormalizable quantum field theories.  Our consideration is
mainly based on Ref. [23]. Finally, we study the dependence of the
low-energy effective action in ${\cal N}=4$ SYM theory on
hypermultiplet fields.

\section{${\cal N}=2$ SYM Theory \\
in ${\cal N}=1$ Superspace}

The simplest and well developed description of four-dimensional
supersymmetric field theories is formulation in terms of ${\cal
N}=1$ superspace. From the point of view of ${\cal N}=1$
supersymmetry, a field content of the pure ${\cal N}=2$ SYM model
is given by the vector multiplet superfield $V$ and chiral
superfield ${\Phi}$, and the field content of the hypermultiplet
is given by two chiral superfields $Q_{+}, \bar{Q}_{-}$. This
allows one to write the action $S$ of the ${\cal N}=2$ SYM model
coupled to hypermultiplet matter in ${\cal N}=1$ superspace as
follows:
\begin{eqnarray}
S&=&S_{\rm SYM} + S_{\rm Hyper}\\ S_{\rm SYM}&=&{1 \over
T(R)g^2}{\rm tr}[\int d^6z\, {1 \over
2}W^{\alpha}W_{\alpha}+\nonumber\\
&+& \int d^8z\, \bar{\Phi}{\rm
e}^{V}\Phi {\rm e}^{-V}], \\ S_{\rm Hyper}&=& \int d^8z
\,(\bar{Q}_{+}{\rm e}^{V}Q_{+}+Q_{-} {\rm e}^{-V}\bar{Q}_{-})+
\nonumber\\
&+& i\int d^6z \, Q_{-}\Phi Q_{+} + i\int d^6\,
\bar{z}\bar{Q}_{+}\bar{\Phi}\bar{Q}_{-},
\end{eqnarray}
where the superfields $V=V^{A}T^{A}$ and $\Phi=\Phi^{A}T^{A}$ form
the ${\cal N}=2$ gauge multiplet with the component fields
$(A_{\mu}, \lambda_+, \phi)$ belonging to the adjoint
representation of the gauge group $G$, and the superfields $Q_+$
form a hypermultiplet with the component fields $(\psi_{+},H_+,
\psi_{-})$ belonging to some representation ${\cal R}$ of $G$. We
use the conventions of Ref. [24].

The classical actions $S_{\rm SYM}$ and $S_{\rm Hyper}$ are gauge
invariant and  manifestly ${\cal N}=1$ supersymmetric by
construction. However, the full action $S$ is also invariant under
the hidden ${\cal N}=2$ supersymmetry transformations, which can
be written in terms of the covariant chiral superfields
$\Phi_{c}={\rm e}^{\bar\Omega}\Phi{\rm e}^{-\bar\Omega}$, $Q_{+
c}= {\rm e}^{\bar\Omega}Q_{+}$ etc.:
\begin{eqnarray}
&\delta\Phi_{c}=\epsilon^{\alpha}W_{\alpha},& \nonumber\\
 &\delta\bar{\Phi}_{c}=\bar{\epsilon}^{\dot{\alpha}}\bar{W}_{\dot{\alpha}},&
\nonumber\\ & \delta
W_{\alpha}=-\epsilon_{\alpha}\bar{\nabla}^{2}\bar{\Phi}_{c}+
i\epsilon^{\dot{\alpha}}\nabla_{\alpha\dot{\alpha}}\Phi_{c}, &
\nonumber\\ &
\delta\bar{W}_{\dot{\alpha}}=-\bar{\epsilon}_{\dot{\alpha}}\nabla^{2}\Phi_{c}+
i\epsilon^{\alpha}\nabla_{\alpha\dot{\alpha}}\bar{\Phi}_{c},&
\end{eqnarray}

\begin{eqnarray}
&\delta\bar{Q}_{+\,c}=\bar{Q}_{+\,c}(\Delta_{1} \Omega)- \nabla^{2}
(Q_{-\,c}\chi),& \nonumber\\
&\delta\bar{Q}_{-\,c}=-(\Delta_{1} \Omega)\bar{Q}_{-\,c}
+\nabla^{2}(\chi Q_{+\,c}),& \nonumber\\
&\delta Q_{+\,c}=- (\Delta_{2} \Omega)Q_{+\,c} +\bar{\nabla}^{2}(\chi
\bar{Q}_{-\,c}), & \nonumber\\
 &\delta Q_{-\,c}=Q_{-\,c}(\Delta_{2} \Omega)-
\bar{\nabla}^{2}(\bar{Q}_{+\,c}\chi),& \nonumber\\ &\Delta_{1}
\Omega = {\rm e}^{-\Omega}\delta {\rm e}^{\Omega}= i\chi\Phi_{c},&
\nonumber\\ & \Delta_{2} \Omega = {\rm e}^{\bar\Omega}\delta {\rm
e}^{-\bar\Omega}= i\bar{\Phi}_{c}\chi,& \nonumber\\ &\chi=
\lambda(\theta)+ \bar{\lambda}(\bar\theta).&
\end{eqnarray}
Here $\Omega$ is a complex superfield determining the gauge
superfield $V$ in the form ${\rm e}^{V}={\rm e}^{\Omega}{\rm
e}^{\bar{\Omega}}$, $\lambda$ and $\bar\lambda$ are chiral and
antichiral space-time-independent superfield parameters with the
expansion $\lambda=\gamma + {1\over
2}\theta^{\alpha}\epsilon_{\alpha}+
\theta^{2}(\beta_{1}+i\beta_{2})$, where the $\beta_{1}$ and
$\beta_{2}$ parameterize the $SU(2)/U(1)$ group,
$\epsilon_{\alpha}$ are the anticommutative parameters present in
the Eqs. (4), and $\gamma$ parameterizes the central charge
transformations. The hypermultiplet action and corresponding
${\cal N}=2$ supersymmetry transformations in terms of ${\cal N}
=1$ superspace were considered in Refs. [24] and [25]. Invariance
of the actions $S_{SYM}$ and $S_{Hyper}$ under the transformations
(4, 5) can be checked directly. One also points out that both the
${\cal N}=2$ super Yang-Mills model and the hypermultiplet model
are the superconformal invariants [26]. Further we will use only
the covariant chiral superfields, and subscript $c$ will be
omitted.

The low-energy effective action of the model under consideration
is described by the holomorphic scale-dependent effective
potential ${\cal F({\cal W})}$ and the non-holomorphic
scale-independent real effective potential ${\cal H}({\cal W},
\bar{\cal W})$ where ${\cal W}$ is ${\cal N}=2$ superfield
strength. The corresponding contributions to the effective action
can be expressed in terms of ${\cal N}=1$ superfields. The
holomorphic part $\Gamma_{F}$ of low-energy effective action is
written in ${\cal N}=1$ form as follows [16]
\begin{eqnarray}
& \Gamma_{F}= \int d^{4}x d^{2}\theta\,
{1\over 2}{\cal F}_{AB}(\Phi)W^{A\alpha} W^{B}_{\alpha} + & \nonumber \\
& +\int d^{4}x d^{4}\theta\,{\cal F}_{A}(\Phi)\bar{\Phi}^{A}+ h.c. &
\end{eqnarray}

The non-holomorphic contribution $\Gamma_{H}$ can be given in an
${\cal N}=1$ form using the metric, connection and curvature of
natural K\"ahler geometry since the ${\cal H}$ is associated with
a K\"ahler potential on a complex manifold defined modulo the real
part of a holomorphic function
\begin{eqnarray}
\Gamma_{H}&=&\int d^{4}x d^{4}\theta\, (g_{A\bar{B}}[-{1\over
2}\nabla^{\alpha\dot{\alpha}}\Phi^{A}
\nabla_{\alpha\dot{\alpha}}\bar{\Phi}^{B}+\nonumber\\
&+&i\bar{W}^{B\dot{\alpha}}(\nabla^{\alpha}_{\dot{\alpha}}W^{A}_{\alpha}+
\Gamma^{A}_{CD}\nabla^{\alpha}_{\dot{\alpha}}\Phi^{C}W^{D}_{\alpha})-\nonumber\\
&-&(f^{ACD}\bar{W}^{B\dot{\alpha}}\Phi^{C}\bar{\nabla}_{\dot{\alpha}}\bar{\Phi}^{D}+\nonumber\\
&+&f^{BCD}W^{A\alpha}\bar{\Phi}^{C}\nabla_{\alpha}\Phi^{D})+\nonumber\\
&+&(\nabla^{2}\Phi^{B}+{1\over
2}\Gamma^{\bar{B}}_{\bar{C}\bar{D}}\bar{W}^{C\dot{\alpha}}
\bar{W}^{D}_{\dot{\alpha}})\times\nonumber\\
&\times&(\bar{\nabla}^{2}\bar{\Phi}^{A}+{1\over 2}\Gamma^{A}_{EF}
W^{E\alpha}W_{\alpha}^{F})]+ \nonumber\\ &+&{1\over
4}R_{A\bar{B}C\bar{D}}(W^{A\alpha}W^{C}_{\alpha}\bar{W}^{B\dot{\alpha}}
\bar{W}^{D}_{\dot{\alpha}})+\nonumber\\ &+& f^{AED}{\cal
H}_{D}\Phi^{E}({1\over 2}\nabla^{\alpha}W^{A}_{\alpha}+
f^{ABC}\Phi^{B}\bar{\Phi}^{C})),
\end{eqnarray}
where
$g_{A\bar{B}}={\cal H}_{A\bar{B}}$,
$\Gamma^{A}_{BC}=g^{A\bar{D}}{\cal H}_{BC\bar{D}}$,
$R_{A\bar{B}C\bar{D}}={\cal
H}_{AC\bar{B}\bar{D}}-g_{E\bar{F}}\Gamma^{E}_{AC}\Gamma^{\bar{F}}_{\bar{B}\bar{D}}$.
Being expressed in terms of component fields, the contribution to effective action
$\Gamma_{H}$ contains at most four space-time derivatives.

We will analyze a general form of the one-loop functionals
$\Gamma_{F}$ and $\Gamma_{H}$ in the model under consideration
using functional methods in the ${\cal N}=1$ superspace and revise
the contributions to the effective action which determine a
functional dependence of ${\cal F}$ and ${\cal H}$ on the ${\cal
N}=2$ vector multiplet. Eqs. (6, 7) play a very important role in
such an approach since they ensure a bridge between the ${\cal N}
=1$ and ${\cal N}=2$ descriptions and allow one to restore
manifestly ${\cal N}=2$ supersymmetric functionals on the basis of
their ${\cal N}=1$ projections.

\section{Background Field Quantization}
The background field method is a powerful and convenient tool for
studying the structure of a quantum gauge theory (see Refs. [24,
27]). After fields splitting, the action in the ${\cal N}=1$ SYM
theory with matter multiplets will be written as a functionals of
the background superfields ${\Omega}, {\bar\Omega}, {\Phi},
{\bar\Phi}$ and quantum ones $V, {\phi}, {\bar\phi}$. To quantize
the theory, we impose the gauge-fixing conditions only on the
quantum fields, introduce the corresponding ghosts and consider
the background fields as functional arguments of the effective
action.

We choose the proper gauge-fixing conditions for the quantum
superfields  $V$ and ${\phi}$ in the form
\begin{eqnarray}
&\bar{F}^{A}=\nabla^2V^{A} +i\lambda {1\over
\Box_{+}}\nabla^{2}\phi^{B}\bar{\Phi}^{C}f^{ABC}, & \nonumber\\
&F^{A}=\bar{\nabla}^2V^{A} -i\bar{\lambda} {1\over
\Box_{-}}\bar{\nabla}^2\bar{\phi}^{B}\Phi^{C}f^{ABC}& ,
\end{eqnarray}
where $\lambda, \bar{\lambda}$ are the arbitrary numerical
parameters and $\Box_\pm$ are standard notations for Laplace-like
operators in the superspace. It is evident that the gauge fixing
functions (8) are covariant under background superfield
transformations. The gauge fixing functions (8) can be considered
as a superfield form of so-called $R_{\xi}$-gauges which are
ordinarily used in spontaneously broken gauge theories. An
extension of $R_{\xi}$-gauge fixing conditions to ${\cal N}=1$
superfield theories has been given in Ref. [28].

The gauge-fixing action corresponding to the functions (8) is
constructed in the standard form
\begin{equation}
S_{\rm GF}=-{1\over \alpha g^{2}}\int
d^8z\,(F^{A}\bar{F}^{A}+b^{A}\bar{b}^{A})
\end{equation}
and depends on the extra parameter ${\alpha}$. The quadratic part
of the Faddeev-Popov action which is relevant for one loop is
\begin{eqnarray}
S_{ghost}&=&\int d^8z\, (\bar{c}^{'A}c^{A}-c^{'A}\bar{c}^{A}+\nonumber\\
&+&\bar{c}^{B}{\lambda\over\Box_{-}}\bar{X}^{BE}X^{EA}c^{'A}+\nonumber\\
&+&\bar{c}^{'B}{\bar{\lambda}\over\Box_{-}}\bar{X}^{BE}X^{EA}c^{A}+b^{A}\bar{b}^{A}),
\end{eqnarray}
where
$X^{AB}= f^{ACB}\Phi^{C},\quad
\bar{X}^{AB}=f^{ACB}\bar{\Phi}^{C}. $

All one-loop contributions to the effective action are given in
terms of the functional trace ${\rm Tr}\ln (\hat{H})$, where the
operator $\hat{H}$ is the matrix of the second variational
derivatives of the action $S_{2}$ in all quantum fields. The
one-loop effective action in the model under consideration reads
\begin{eqnarray}
\Gamma [V,\Phi]&=&{i\over 2}{\rm Tr}\ln \hat{H}_{\rm SYM}+i{\rm
Tr}\ln \hat{H}_{\rm Hyper}-\nonumber\\
&-& {i\over 2}{\rm Tr}\ln \hat{H}_{ghost}.
\end{eqnarray}
It should be noted that operator $\hat{H}_{SYM}$ contains the
contributions from ${\cal N}=1$ vector and chiral multiplets
forming ${\cal N}=2$ gauge multiplet. The choice
$\lambda=\bar\lambda=\alpha$ in (8, 9) greatly simplifies all
calculation because it diagonalizes the matrix $\hat{H}_{SYM}$ and
decoupled the contributions from the ${\cal N}=1$ vector and
chiral multiplets. However, we will keep the gauge parameters
$\lambda$ and $\alpha$ arbitrary and investigate the dependence of
the effective action on these parameters.

\section{${\cal N}=1$ K\"ahler and \\
Non-holomorphic ${\cal N}=2$ \\
Potentials }
\subsection{${\cal N}=1$ K\"ahler potential}

In this section we study the form of the non-Abelian low-energy
effective action $\Gamma=\int d^{8}z\, K$ and its gauge
dependence. It is known that in the non-Abelian case the K\"ahler
potential cannot be written in the form $Im (\bar{\Phi}{\cal
F}'(\Phi))$ consistent with the rigid version of special geometry
(see e.g. Refs. [5, 14]). The additional terms originate from a
real function ${\cal H}_{0}({\cal W},\bar{\cal W})$ of the ${\cal
N}=2$ YM superfield strength ${\cal W}$. The results obtained in
the present paper are more general as compared with the ones
obtained in Refs. [5, 12, 14] since here we have used here the
more general and complicated gauges. To calculate these
potentials, we consider the diagrams with external
$\Phi,\bar{\Phi}$ lines corresponding only to the constant field
background.  Such a choice of background superfields leads to a
number of technical simplifications due to the absence of the
background gauge field, which allows us to replace all background
covariant derivatives with flat ones (i.e. $\nabla\rightarrow D,
\bar{\nabla}\rightarrow \bar{D}$). This provides a possibility of
using the superspace projectors $P_{1} = {1 \over \Box}\bar{D}^2
D^2, \quad P_{2} = {1 \over \Box}D^2 \bar{D}^2, \quad P_{T}= -{1
\over \Box}D \bar{D}^2 D$ and $\Pi_{0}= P_{1}+P_{2}$ and
simplifying the evaluations of the functional determinants (11).

The final result is a sum of three terms:\\
1) The hypermultiplet contribution to effective action
\begin{eqnarray}
K_{\rm Hyper}^{fund} = {-1\over (8\pi)^{2}}(\Phi\bar{\Phi})\left(
\ln {\Phi^{2}\bar{\Phi}^{2}\over 16{\rm e}^{2}\Lambda^{4}}+ s\ln
{1+s\over 1-s}\right),
\end{eqnarray}
where $\bar{\Phi}\Phi$ denotes the scalar product in isospin
space, and we have used the notation $ s^2 =
1-{\Phi^{2}\bar{\Phi}^{2}\over (\Phi\bar{\Phi})^{2}} < 0.
$\\
2) The effective action $\Gamma_{\rm SYM}=\Gamma_{\rm V}+
\Gamma_{\rm GD}$ induced by the ${\cal N}=2$ vector multiplet
contains the vector loop contribution
\begin{eqnarray}
&K_{\rm V} = {\displaystyle {1\over
(4\pi)^2}}\left((\Phi\bar{\Phi})\ln {\displaystyle
{\Phi^{2}\bar{\Phi}^{2}\over {\rm e}^{2}\Lambda^{4}}}+
(\Phi\bar{\Phi})\ln t \right.&
\end{eqnarray}
$$\left.+\sqrt{\,\Phi^{2}\bar{\Phi}^{2}}\left[{\displaystyle
{t+1\over 2}\ln {t+1\over 2} + {t-1\over 2}\ln {t-1\over 2}}
\right] \right),$$
where the notation $t={{\Phi}\bar{\Phi}\over
\sqrt{{\Phi}^{2}\bar{\Phi}^{2}}}$ was introduced, plus\\
3) the gauge dependent contribution $$ K_{\rm GD}=\int {dk^2\over
(4\pi)^2} \ln \left(1+ {((\bar{\Phi}\Phi)^2
-\Phi^2\bar{\Phi}^2)\over (k^2+\lambda
(\bar{\Phi}\Phi))(k^2+\bar{\lambda}(\bar{\Phi}\Phi))}\right.
$$
$$
\left.\left[-{\lambda\bar{\lambda}\over 2}+\alpha
\left({\lambda\bar{\lambda}\over 4k^2 }(\bar{\Phi}\Phi)+
{\lambda+\bar{\lambda}\over 2}-{\alpha\over 4 } \right)
\right]\right)=
$$
\begin{eqnarray}
=\int {dk^2\over (4\pi)^2} \ln
\left({ (k^{2} - e_{1})(k^2 - e_{2})(k^2 - e_{3})\over
k^{2}(k^{2}+1)^2} \right),
\end{eqnarray}
which automatically vanishes for Abelian background fields $\Phi$.
This is the main result of the subsection. The dependence of the
one-loop effective action on all gauge parameters is given by this
expression. When $\lambda=\bar{\lambda}=0$ the result (12, 13, 14)
coincides with one given in Ref. [12]. The case
$\lambda=0,\,\alpha=1$ is known as the Fermi gauge. The
corresponding form of the K\"ahlerian potential (12, 13, 14) was
found in Ref. [14]. Result for Landau-DeWitt gauge is obtained
with $\alpha=0,\, \lambda=\bar{\lambda}=1$. Note that (14) in the
gauge $\alpha=\lambda=\bar{\lambda}=1$, which can be naturally
called the Fermi-DeWitt, two last terms in the first line in (13)
are exactly cancelled by (14) while the first term (13) is
doubled.

Using the simple integral in (14) we obtain
\begin{equation}
K_{\rm GD}=e_{1}\ln (-e_{1})+ e_{2}\ln (-e_{2})+
e_{2}\ln (-e_{2}),
\end{equation}
where $e$'s are the roots of the polynomial of $k^{2}$ under the
logarithm in (14).

\subsection{${\cal N}=2$ non-holomorphic potential}
In previous subsection we have found the one-loop K\"ahler
effective potential $K(\Phi,\bar{\Phi})$ induced by both ${\cal
N}=2$ vector multiplets and hypermultiplets. As has been mentioned
in Refs. [5, 14], the K\"ahler potential in the non-Abelian case
determines not only by the holomorphic function ${\cal F}$.
Additional terms originate from a real function ${\cal H}({\cal
W}, \bar{\cal W})$ of the ${\cal N}=2$ Yang-Mills superfield
strength ${\cal W}$, which is integrated over full ${\cal N}=2$
superspace. Comparison the last term in decomposition (7) with the
K\"ahler potential leads to
\begin{eqnarray}
K(\Phi,\bar{\Phi})&=&\bar{\Phi}^{A}{\cal
F}_{A}+\Phi^{2}(\bar{\Phi}^{A}{\cal H}_{A}) -\nonumber\\
&-& (\Phi\bar\Phi)(\Phi^{A}{\cal H}_{A}),
\end{eqnarray}
where
$$
\Phi^{A}{\cal H}_{A}=0,\quad \bar{\Phi}^{A}{\cal H}_{A}=
-{2\bar{\Phi}^2\over (\Phi\bar{\Phi})}s^{2}{\partial {\cal H}\over\partial s^{2}}.
$$

It is well known that the $\beta$-function and the axial anomaly
arise exactly from the holomorphic potential ${\cal F}$. This fact
gives us a unique recipe for extracting the contributions from the
K\"ahler potential, which can be associated with holomorphic and
non-holomorphic potentials respectively. Using the expressions
(12) and (13) and the reconstruction formula (16), ones find, in
accordance with Ref. [5], the contributions to the holomorphic
potential ${\cal F}({\cal W})$ and to the non-holomorphic
potential ${\cal H}({\cal W}, \bar{\cal W})$ depending on the
${\cal N}=2$ superfield strength ${\cal W}$:
\begin{equation}
{\cal F}_{\rm Hyper}^{fund} = {-1\over (8\pi)^2}{\cal W}^2\ln
{{\cal W}^2 \over {\rm e}^2 \Lambda^2},
\end{equation}
\begin{equation}
{\cal F}_{\rm Vector} = {1\over (4\pi)^2}{\cal W}^2\ln {{\cal W}^2
\over {\rm e}^2 \Lambda^2},
\end{equation}
\begin{equation}
{\cal H}_{\rm Hyper}= {1\over (16\pi)^2} \ln^2
{1+s \over 1-s},
\end{equation}

\begin{equation}
{\cal H}_{\rm Vector} = {1\over (8\pi)^2}\left(  {\rm
Li}_{2}(1-t^{2}) - 2\ln {t+1\over 2}\ln {t-1 \over 2}\right)
\end{equation}

Our further aim is to obtain the off-shell gauge-dependent
contribution to ${\cal H}$ from the gauge-depen\-dent part of the
full K\"ahler potential. In this case Eq. (16) is written in the
form
\begin{equation}
-2s^{2}(1-s^{2}){d{\cal H}_{\rm
GD}\over ds^{2}}= \tilde{K}_{\rm GD}(s^{2}),
\end{equation}
where $\tilde{K}_{\rm GD}$ was introduced in Eq.(14),
$s^2=1-1/t^{2}$ and $t={{\cal W}\bar{\cal W}\over \sqrt{{\cal
W}^{2}\bar{\cal W}^{2}}}$, $t\in [0,1]$.  It has already been
noticed that $K_{\rm GD} =0$ at $s^{2}\rightarrow 0$ and therefore
${\cal H}_{\rm GD}$ vanishes on-shell.

We see the holomorphic potential ${\cal F}$ is gauge independent.
The whole dependence on the gauge-fixing parameters is
concentrated in the term ${\cal H}_{\rm GD}$ of the
non-holomorphic potential ${\cal H}$. Let us present (14) as a
formal power series.  Eq. (15) is nothing but a determination of a
symmetrical function via the polynomial roots. According to the
fundamental theorem in theory of symmetrical functions (see e.g.
Ref. [29]) "any entire rational symmetrical function can be
uniquely rewritten as a entire rational function of elementary
symmetrical functions" (i.e. coefficients of the polynomial). To
represent (15) as an entire rational function, we expand the
logarithms into a formal power series
\begin{equation}
K_{\rm GD} =
-\sum_{n=1}^{\infty}{1\over n}\,S_{n},
\end{equation}
where the {\it power}
symmetrical functions of the roots $e_{1}, e_{2}, e_{3}$ of the
form
\begin{equation}
S_{n} = e_{1}(1 + e_{1})^{n} + e_{2}(1 +
e_{2})^{n} + e_{3}(1 + e_{3})^{n}
\end{equation}
has been used. Using Newton's classical recursion formulae, we can
uniquely express $S_{n}$ in terms of elementary symmetrical
functions. It is well known that the roots $e_{i}$ of an algebraic
equation are always satisfy the Vieta relations. For the roots of
the polynomial, which appears from the numerator in the logarithm
of (14) we have
\begin{equation}
 -e_{1}e_{2}e_{3}=g_{3},\quad e_{1}e_{2} +
e_{2}e_{3} + e_{1}e_{3} = g_{2}, \end{equation}$$\quad e_{1} +
e_{2} + e_{3} = -2,$$ where elementary symmetrical functions are
given from (14, 24) as $ g_{2}=1+s^2(-{1\over
2}+\gamma(1-{\gamma\over 4})),\quad g_{3}=s^2{\gamma\over 4}.$
Multiplying Eq. (23) by $e_{1} + e_{2} + e_{3}$ and using
identities (24) we obtain the recursion relation
\begin{equation}
S_{n+1} - S_{n} - (1 - g_{2})S_{n-1}+(1-g_{2}+g_{3})S_{n-2} = 0.
\end{equation}
Using this relation, one can evaluate any $S_{n}$ step by step.
One can check that $S_{n}\sim s^{2}$ for any $n$, and each $S_{n}$
includes $g_{3}$ linearly. It allows one to simplify integration
in Eq. (21).

Moreover, Waring's well-known formulae (see. e.g. [29]) allow to
express $S_{n}$ for any $n$ directly in terms of $g_{2},\,g_{3}$.
In order to get all $S_{n}$, it is very useful to introduce a
generating function defined by a formal power series
\begin{equation}
G(\tau) =\sum_{k=1}^{\infty}\,\tau^{k-1}S_{k},
\end{equation}
then any $S_{n}$ can be found with help of differentiations of the
generating function $G$ with respect to $\tau$. It also allows us
to express the general term of the sequence $S_{n}$ in terms of
symmetrical functions $g_{2}$ and $g_{3}$ instead of the roots
$e_{i}$. Since the functions $g_{2},\,g_{3}$ are known from the
integral (14), we can avoid finding the roots $e_{i}$ for analysis
${\cal H}_{\rm GD}$ at all. The generating function $G$ satisfies
an algebraic equation which can be derived by multiplying the
recursion relation by $\tau^{k}$ and summing over powers $k$. The
solution to this equation is
\begin{equation} G(\tau) = {2(1 - g_{2})(1-2\tau + \tau^{2}) -
3g_{3}\tau +2g_{3}\tau^{2} \over 1 - \tau - (1 - g_{2})\tau^{2} +
(1 - g_{2} + g_{3})\tau^{3}}.
\end{equation}
As a result we obtain an expansion of
$K_{\rm GD}$ in terms of elementary symmetrical functions $g_{2}$,
$g_{3}$:
\begin{equation}
K_{\rm GD} =
-\sum_{n=1}^{\infty}{1\over n!} \left({d\over d\tau}\right)^{n-1}
G(\tau)|_{\tau=0}.
\end{equation}
Now, it is useful to introduce the new parameters
$g=-1/2+(\gamma/2-1)^{2},\, g_{3}=\gamma/4,\,p=g+g_{3}$,
$u=1-s^{2}$. Using the binomial formula for derivatives of the
generating function (27) in (28), we rewrite the equation (21) in
the following form
\begin{equation}
-2u{d{\cal H}_{\rm GD}\over du}=\sum_{k=0}^{\infty}{1\over (k+3)!}\times
\end{equation}
$$
\times
\left(4g-g_{3}(k+1)(k+5)\right)\left({d\over d\tau}\right)^{k}Y|_{\tau=0},
$$ where
$ Y^{-1}=1-\tau-g\tau^2+p\tau^3+u(g\tau^2-p\tau^3). $ It is useful
to extract, in the right hand side of Eq. (29), the powers of $u$
and to rewrite this relation in form of double sum. This
expression allows one to find ${\cal H}_{\rm GD}$ as a series with
a coefficients of each given power of $u$ depending on elementary
symmetrical functions. Hence, we finally can rewrite (21) in terms
of elementary symmetrical functions. We see that the right hand
side (29) can be written via rational functions for any given
choice of gauge parameters. For some partial choice of gauge
parameters, arbitrary term of series can be found exactly.

Let's consider the Landau-DeWitt gauge in more detail. At such a
choice, $Y^{(k)}$ in (29) becomes simple enough
\begin{equation}
Y^{(k)}=k!\left({1\over 1-a^{2}}-{(-a)^{k+1}\over
2(1+a)}-{a^{k+1}\over 2(1-a)} \right),
\end{equation}
$$\quad a^{2}={s^{2}\over 2}$$ and the general term in right side
(29) can be exactly found. Finally, Eq. (29) becomes
\begin{equation}
(1-2a^{2}){d{\cal H}_{\rm GD}\over da}={1-a\over
a}\ln (1-a)+{1+a\over a}\ln (1+a)
\end{equation}
and we obtain ${\cal H}_{\rm
GD}$ by integration
\begin{eqnarray}
2(4\pi)^{2}{\cal H}_{\rm GD}&=& \ln (2)\ln (1-s^{2})+
\end{eqnarray}
$$
+{1\over \sqrt{2}}\ln \left({\sqrt{2}-1\over \sqrt{2}+1}\right)
\ln (1-s^{2})-{\rm Li}_{2}\left({s^{2}\over 2}\right)+
$$
$$
+{\sqrt{2}-1\over\sqrt{2}} \left[{\rm
Li}_{2}\left({s-1\over\sqrt{2}-1}\right)+ {\rm
Li}_{2}\left(-{s+1\over\sqrt{2}-1}\right)\right]+
$$
$$
+{\sqrt{2}+1\over\sqrt{2}} \left[{\rm
Li}_{2}\left({s+1\over\sqrt{2}+1}\right)+ {\rm
Li}_{2}\left({1-s\over\sqrt{2}+1}\right)\right].
$$

We emphasize that expressions (19), (20) and (32) are exact
results within the one-loop approximation. Of course, they can be
expanded in series in two limit cases: $t\rightarrow 1$ and
$t\rightarrow 0$. Such a behavior is not unusual and it looks like
quite similar to the well-known exactly solvable model in an
effective field theory, namely the Euler-Heisenberg effective
action. One can point out some more property of the
Euler-Heisenberg effective action at small mass (strong external
field): it possesses by logarithmic branch point as well as ${\cal
H}_{\rm GD}$, ${\cal H}_{\rm Vector}$, while at large mass (weak
external field) there exists an asymptotic series expansion in
inverse powers of mass.

We have shown that the gauge-dependent part of the off-shell
effective action can be found with an arbitrary level of accuracy
and at any choice of the gauge fixing parameters. The form of the
non-holomorphic effective potential has an essential arbitrariness
due to its explicit gauge dependence. In particular, this fact
leads to the ambiguous definition of
$$R_{A\bar{B}C\bar{D}}(W^{A\alpha}W^{C}_{\alpha}\bar{W}^{B\dot{\alpha}}
\bar{W}^{D}_{\dot{\alpha}})$$ term from Eq. (7), which should
reproduce the leading term in the expansion of the non-Abelian
analog of the Born-Infeld action (see, e.g. [21]). The structure
of the tensor $R_{A\bar{B}C\bar{D}}$ is cumbersome enough. In
addition, we point out that the symmetrized trace
$(F^{+})^{2}(F^{-})^2/\phi^{2}\bar{\phi}^{2}$, determining the
full set of $F^{4}$-terms in the effective action, also contains
the various contractions $\phi^{A},\,\bar{\phi}^{A}$ with $F^{A}$.
The existence a large class of gauge theory operators, which
correspond to supergravity modes and contain nontrivial extra
factors (depending on $\phi^{A},\,\bar{\phi}^{A}$), in the
non-Abelian Born-Infeld action was discussed in Ref. [21].

To conclude this subsection, we note that, unlike the Abelian
case, ${\cal N}=2$ supersymmetry itself can not uniquely fix a
form of next-to-leading term in the effective action because of
its explicit gauge dependence.

\section{Complete ${\cal N}=4$ structure \\
of the low-energy effective action
in ${\cal N}=4$ SYM theories }

The ${\cal N}=4$ SYN theory, being maximally extended rigid
supersymmetric model, possesses the remarkable properties on
classical and quantum levels. The corresponding quantum theory is
finite, scale independent and superconformally invariant. The
exact low-energy quantum dynamics of this model is described by a
non-holomorphic effective potential [6-9]. The explicit form of
the non-holomorphic potential for the $SU(N)$ gauge group
spontaneously broken down to $U(1)^{N-1}$ is given by the
expression
\begin{equation}
{\cal H}({\cal W}, \bar{\cal W}) = c\sum_{I<J}\, \ln \left({{\cal
W}^{I}-{\cal W}^{J}\over \Lambda}\right)\times
\end{equation}
$$ \times\ln \left({\bar{\cal W}^{I}-\bar{\cal W}^{J}\over
\Lambda}\right),
$$
where  $\Lambda$ is an scale and $c =1/(4{\pi})^2$ (see e.g. [8]).
Expression (33) determines exact low-energy effective action in
the ${\cal N}=2$ gauge superfield sector.

We point out that the result (33) is so general that it can be
obtained entirely on the symmetry grounds, from the requirements
of scale independence and $R$-invariance only up to a numerical
factor [6]. Moreover, the potential (33) gets neither quantum
corrections beyond one loop nor instanton corrections [6]. These
properties are very important for understanding low-energy quantum
dynamics in ${\cal N}=4$ SYM theory in the Coulomb phase. In
particular, this effective potential provides description of
sub-leading terms in the interaction between parallel D3-branes in
superstring theory [21].

The complete exact low-energy effective action containing the
dependence on both the ${\cal N}=2$ gauge superfields and the
hypermultiplets has been discovered [31] using a techniques of
harmonic superspace [33]. It was shown that the algebraic
restrictions on the full ${\cal N}=4$ supersymmetric structure of
the low-energy effective action are so strong that they allows us
to restore the dependence of effective action on the
hypermultiplets on the basis of the known non-holomorphic
effective potential (33). As a result, the additional to (33)
hypermultiplet dependent contribution, containing the on-shell
${\cal W}$, $\bar{\cal W}$ and hypermultiplet $q^{+a}$ [32]
superfields, has been obtained in the form
\begin{equation}
{\cal L}_{q}=c\left\{(X-1)\frac{\ln
(1-X)}{X}+[Li_{2}(X)-1]\right\}, \quad
\end{equation}
$$
X=-\frac{q^{ia}q_{ia}}{{\cal W}\bar{\cal W}}~.
$$
The effective Lagrangian (34) together with the effective potential (33) define
the exact ${\cal N}=4$ supersymmetric low-energy effective action
in the theory under consideration.

The effective Lagrangian (34) has been found in Ref. [31] on the
purely algebraic ground. It would be extremely interesting to
derive this Lagrangian in the framework of quantum field theory.
Here we just present such a derivation. To be more precise, we
discuss the calculations of the one-loop effective action
depending on both the ${\cal N}=2$ gauge and the hypermultiplet
background fields using the formulation of the ${\cal N}=4$ SYM
theory in terms of ${\cal N}=1$ superfields [24, 27] and exploring
the derivative expansion technique in ${\cal N}=1$ superspace
[15]. It allows us to obtain the exact coefficients by various
powers of covariant derivatives on a constant space-time
background belonging to the Cartan subalgebra of the gauge group
$SU(n)$
\begin{equation}
{\cal W}|=\Phi=Const, \quad D^{i}_{\alpha}{\cal
W}|=\lambda^{i}_{\alpha}=Const,
\end{equation}
$$ D^{i}_{(\alpha}D_{\beta )i}{\cal W}|=f_{\alpha\beta}=Const,\quad
D^{\alpha (i}D^{j)}_{\alpha}{\cal W}|=0,
$$
where $\Phi^{I}=diag(\Phi^{1},\Phi^{2},\ldots,\Phi^{n}),$
$\sum\Phi^{I}=0.$ Another approach to derivations of the
Lagrangian (34) can be developed in ${\cal N}=2$ harmonic
superspace [32]. The structure of the one- and two-loop low-energy
effective actions in ${\cal N}=2$ supersymmetric models is also
studied in Ref. [35].

The main technical feature used in the given paper consists in
the background covariant gauge-fixing multi-parametrical
conditions (8). Since the Abelian background is a solution of the
equations of motion, we won't worry about the choice of the
gauge-fixing parameters. It is therefore convenient to choose the
gauge-fixing which earlier was named the Fermi-DeWitt gauge:
$\alpha=\lambda=1$. The choice of the gauge parameters allows us
to avoid the calculation problems with mixed loops containing
vector-chiral superfield propagators circulating along the loops.

The ${\cal N}=4$ "on-shell" multiplet can be obtained by combining
three ${\cal N}=1$ chiral superfields and one ${\cal N}=1$ vector
superfield [24]. In this description an $SU(3)\otimes U(1)$
subgroup of the $SU(4)$ $R$-symmetry group is manifest. The form
of the ${\cal N}=1$ sypersymmetric action in the chiral
representation  is given by
\begin{eqnarray}
S &=& {1\over g^{2}}{\rm tr}\{\int d^{4}x
d^{2}\theta\,W^{2}+\nonumber\\
&+& \int d^{4}x d^{4}\theta\,\bar{\Phi}_{i}{\rm
e}^{V}\Phi^{i}{\rm e}^{-V}+\nonumber\\
&+& {1\over 3!}\int
d^{4}x d^{2}\theta\, ic_{ijk}\Phi^{i}[\Phi^{j},\Phi^{k}]+\nonumber\\
&+&{1\over
3!}\int d^{4}x d^{2}\bar\theta\,
ic^{ijk}\bar\Phi_{i}[\bar\Phi_{j},\bar\Phi_{k}] \}
\end{eqnarray}
It is convenient to introduce the new notations
$\Phi^{1}=\Phi,\,\Phi^{2}=Q,\,\Phi^{3}=\tilde{Q}$ and to rewrite the two
last terms in (36) as follows
$$
i\int d^{4}x d^{2}\theta\,
Q[{\Phi},\tilde{Q}]+i\int d^{4}x d^{2}\bar\theta\,
\bar{Q}[\bar\Phi,\bar{\tilde{Q}}].
$$
After splitting each field
into the background and quantum parts (i.e. ${\rm e}^{V_{tot}}={\rm
e}^{\Omega}{\rm e}^{gV}{\rm e}^{\bar\Omega}$,$\Phi\rightarrow \Phi
+ \varphi,\,\bar{\Phi}\rightarrow \bar{\Phi} + \bar{\varphi},\,
Q\rightarrow Q + q,\,\tilde{Q}\rightarrow \tilde{Q} + \tilde{q},\,
\bar{Q}\rightarrow \bar{Q} + \bar{q},\, \bar{\tilde{Q}}\rightarrow
\bar{\tilde{Q}} + \bar{\tilde{q}}$) we can rewrite the quadratic part
of the classical action (36), and (9) in a form which does not
contain any $V\Phi$ terms
\begin{equation}
S_{(2)}= -{1\over2}\sum_{I<J}\int d^{4}x d^{4}\theta\, ({\cal
F}{\bf H}{\cal F^{\dag}}+
\end{equation}
$$
+ \bar{V}(O_{V}- M) V),
$$
where $ {\cal F}= (\bar{\varphi},
{\varphi}, \bar{q}, q, \bar{\tilde{q}}, \tilde{q}), \quad {\cal F
}^{\dag}=({\varphi}, \bar{\varphi}, q, \bar{q}, \tilde{q},
\bar{\tilde{q}} )^{T},$
\begin{equation} M_{IJ}=
(\bar{\Phi}_{IJ}\Phi_{IJ}+\bar{Q}_{IJ}Q_{IJ}+\bar{\tilde{Q}}_{IJ}\tilde{Q}_{IJ}),
\end{equation}
$$
O_{V}=\Box-iW^{\alpha}_{IJ}\nabla_{\alpha}-i\bar{W}^{\dot{\alpha}}_{IJ}\bar{\nabla}_{\dot{\alpha}},
$$
where $W^{\alpha}_{IJ}= W^{\alpha}_{I}-W^{\alpha}_{J},$ $
\bar{W}^{\dot{\alpha}}_{IJ}=\bar{W}^{\dot{\alpha}}_{I}-\bar{W}^{\dot{\alpha}}_{J}$
are the background field strengths and $\quad \Phi_{IJ}=
\Phi_{I}-\Phi_{J},\ldots$ Here ${\bf H}$ denotes some $6\times6$
matrix depending on covariant derivatives and background fields.
The explicit form of this matrix and the details of the
calculation are given in Ref. [34].

According to the Faddev-Popov procedure, we also need to introduce
a gauge-compensating term (10) in the action. The final step is
integration in the functional integral over all quantum
superfields. It allows to write the standard representation for
the one-loop effective action
\begin{equation}{\rm
e}^{i\Gamma}=\prod_{I<J}{\rm Det}^{-1} (O_{V}-M){\rm Det}^{-1} ({\bf
H}){\rm Det}^{2} ({\bf H}_{FP})
\end{equation}
Calculation of the functional trace leads to
\begin{eqnarray}
\Gamma_{SYM}&=&i{\rm Tr}\ln (O_{V}-M)+\\ &+&2i{\rm Tr}\left(\ln
(1-{M\over \Box_{+}}){\nabla^{2}\bar\nabla^{2}\over
\Box_{+}}\right)+\nonumber\\ &+&2i{\rm Tr}\left(\ln (1-{M\over
\Box_{-}}){\bar\nabla^{2}\nabla^{2}\over
\Box_{-}}\right),\nonumber
\end{eqnarray}
where $\Box_\pm$ are standard notation for
$\nabla^{2}\bar{\nabla}^{2}$ and $\bar{\nabla}^{2}\nabla^{2}$. In
the space of chiral and antichiral superfields these operators act
as follows
$$
\nabla^{2}\bar{\nabla}^{2}= \Box_{+}=
\Box-i\bar{W}^{\dot{\alpha}}\bar{\nabla}_{\dot{\alpha}}- {i\over
2}(\bar{\nabla}\bar{W}),
$$
$$
\quad \bar{\nabla}^{2}\nabla^{2}=
\Box_{-}= \Box-i W^{\alpha}\nabla_{\alpha}-{i\over 2}(\nabla W).
$$
Evaluation of ${\rm Tr}\ln(H_{FP})$ leads to the following
ghost contribution to the effective action
\begin{eqnarray}
\Gamma_{FP}&=& -2i\left({\rm Tr}\ln (1-{M\over \Box_{-}}){1\over
\Box_{-}}\bar{\nabla}^{2}\nabla^{2}\right)-\nonumber\\
&-&2i\left({\rm Tr}\ln(1-{M\over \Box_{+}}){1\over
\Box_{+}}\nabla^{2}\bar{\nabla}^{2}\right),
\end{eqnarray}
which exactly cancels the second and third terms in (39). This
surprising cancellation between the ghost and chiral fields
contributions in the ${\cal N}=4$ SYM theory effective action was
firstly noted in [8].

After the functional trace calculation, the first term in (39) gives known result [15,26]
\begin{equation}
\Gamma = \frac{1}{8\pi^{2}}\int{\rm d}^8 z \int_{0}^{\infty}{\rm
d}t\,t\,{\rm e}^{-t}\frac{ W^{2}\bar{W}^{2}}{M^2}\,\omega (t\Psi,
t\bar{\Psi}),
\end{equation}
$$ \omega (t\Psi, t\bar{\Psi})=
\frac{\cosh(t\Psi)-1}{t^{2}\Psi^{2}}\frac{\cosh(t\bar\Psi)
-1}{t^{2}{\bar \Psi}^{2}}\times $$ $$ \times\frac{t^{2}(\Psi^{2} -
{\bar\Psi}^{2})} {\cosh(t\Psi) - \cosh(t\bar\Psi)}, $$ $\Psi$ and
$\bar \Psi$ are given by
\begin{equation}
{\bar \Psi}^2 = \frac{1}{M^2} \,\nabla^2 {W^2}, \qquad \Psi^2 =
\frac{1}{M^2} \,{\bar \nabla}^2 {\bar W}^2 .
\end{equation}
One can show that the quantity ${M}=
(\bar{\Phi}\Phi+\bar{Q}Q+\bar{\tilde{Q}}\tilde{Q})$ denominators (42, 43) is
invariant under the $R$-symmetry group of ${\cal N}=4$ supersymmetry.
The function $\omega$ introduced in (42) has the following
expansion
\begin{equation}
\omega (x,y)= \frac{1}{2} +
\frac{x^{2}y^{2}}{4\cdot5!} - \frac{5}{12\cdot7!}\,(x^{4}y^{2}
+x^{2}y^{4})
\end{equation}
$$ + \frac{1}{34500\,}(x^{2}y^{6} + x^{6}y^{2}) +
\frac{1}{86400}\,x^{4}y^{4} + \ldots
$$
Eq. (44) allows one to expand
the effective action (42) in series in powers $\Psi^2$,
$\bar{\Psi}^2$ as follows
\begin{equation} \Gamma =
\Gamma_{(0)}+\Gamma_{(2)}+\Gamma_{(3)}+\cdots,
\end{equation}
where the term $\Gamma_{(n)}$ contains terms
$c_{m,l}\Psi^{2m}\bar{\Psi}^{2l}$ with $m + l= n$.

The effective action (42) and its expansion (45) are given in terms of
the ${\cal N}=1$ superfields. Our next aim is to find a manifest ${\cal
N}=2$ form of each term in the expansion (45). For this purpose we
extract from $M$ ${\cal N}=1$ form of the $X =
-\frac{\bar{Q}Q+\bar{\tilde{Q}}\tilde{Q}}{\bar{\Phi}\Phi}$, which
was defined in Eq. (34) by $M=\Phi \bar\Phi (1- X)$, and then
expand the denominator $(1/M)^{k}$ from (42) in a power series over $X$.
This expansion of $(1/M)^{k}$ and the reconstruction
expressions
\begin{eqnarray}
\nabla^{4}\ln {\cal W}| &=&
\nabla^{2}\left({W^{\alpha}W_{\alpha}\over
\Phi^{2}}\right)+\ldots,\nonumber\\
(\nabla^{2}_{2}){1\over {\cal W}^{2m}}| &=& {2m
(2m+1)\over\Phi^{2m}}{W^{\alpha}W_{\alpha}\over
\Phi^{2}}+\ldots,
\end{eqnarray}
allow us to obtain the first term in (45)(which is $\sim F^4$)
\begin{equation}
\Gamma_{(0)}={1\over (4\pi)^{2}}\int d^{12}z (\ln {\cal W}\ln
\bar{\cal W}+ \sum_{k=1}^{\infty}{1\over k^{2}(k+1)} X^{k}),
\end{equation}
where $X=\left(-{q^{ia}\bar{q}_{ia}\over {\cal W}\bar{\cal
W}}\right)^{n}$ was defined in (34). The second term in (47) can
be transformed to the form (34) using the power series for Euler's
dilogarithm function and we see that this term is just the
effective Lagrangian (34) found in [31, 32].

The following terms in the expansion (45) can be calculated using
expansion $(1/M)^{k}$ in $X$. Their ${\cal N}=2$ form is
reconstructed by taking into account (46). Using the same analysis,
ones get the term $\Gamma_{(2)}$ in (45) in the form
\begin{equation}
\Gamma_{(2)}= {1\over 2(4\pi)^{2}}\int d^{12}z
{\bf\Psi^{2}\bar{\Psi}^{2}} (1  +
\end{equation}
$$ + {1\over 5!}\sum_{k=1}^{\infty}{(k+5)(k+4)(k+1)\over
(k+3)(k+2)}X^{k}).
$$
Its $X$-independent part was given in [26].
Here the ${\cal N}=2$ chiral combinations ${\bf\bar\Psi^2} = \bar{\cal
W}^{-2}\nabla^{4}\ln{\cal W}$, ${\bf\Psi^2} = {\cal
W}^{-2}\bar\nabla^{4}\ln\bar{\cal W}$ are the scalars under ${\cal
N}=2$ superconformal group transformations. The sum in (48) can be
transformed as follows
$$
 \sum_{k=1}^{\infty}{(k+5)(k+4)(k+1)\over
(k+3)(k+2)}X^{k}= $$ $$={1\over (1-X)^{2}}+{4\over (1-X)}+ $$
$$+{6X-4\over X^{3}}\ln (1-X)-4{X-1\over X^{2}}-{10\over 3}.
$$
Applying the same procedure for the third term in (45) one obtains
\begin{equation}
\Gamma_{(3)}=-{5\over 6\,(4\pi)^{2}}\int
d^{12}z ({\bf\Psi^{4}\bar{\Psi}^{2}+\Psi^{2}\bar{\Psi}^{4}})\times
\end{equation}
$$ \times {1\over 7 !}\sum_{k=1}^{\infty}(k+7)(k+6)(k+1)X^{k},
$$
where the sum in right hand side is
$$
\sum_{k=1}^{\infty}(k+7)(k+6)(k+1)X^{k}= $$ $$ ={2X\over
(1-X)^{4}}(56-116X+84X^{2}-21X^{3}).
$$
Thus, we have found the
hypermultiplet dependence of the contributions $\Gamma_{(0)}$,
$\Gamma_{(2)}$ and $\Gamma_{(3)}$ to the known effective action
[26] which depend on ${\cal N}=2$ vector multiplet. As result we
obtained the complete ${\cal N}=4$ supersymmetric forms for the
three first terms of expansion of the effective action (42) in
power series in Abelian strength. It is evident, that such a
reconstruction procedure can be applied to any term in the expansion
(45).

The fourth term in (45) contains two parts. The first one is
$$
\Gamma_{(4_{1})}={1\over (4\pi)^{2}}{1\over 17250}\int d^{12}z
({\bf \Psi^{2}\bar{\Psi}^{6}+\Psi^{6}\bar{\Psi}^{2}})\times $$ $$
\times{12X\over(1-X)^{6}} (450-1545X+2284X^{2}- $$ $$
-1779X^{3}+720X^{4}-120X^{5})
$$
and the second part is given as follows
$$ \Gamma_{(4_{2})}={1\over 5\cdot 6!}{1\over
(4\pi)^{2}}\int d^{12}z {\bf \Psi^{4}\bar{\Psi}^{4}}\times $$ $$
\times( {12(5X-4)\over X^{5}}\ln (1-X)- $$ $$ -{1\over 5
X^{4}(1-X)^{6}}(240-1620X+4610X^{2}- $$ $$
-7120X^{3}+6363X^{4}-4878X^{5}+ 6135X^{6}- $$ $$
-7560X^{7}+5670X^{8}-2268X^{9}+378X^{10})) $$

As a result, we see that the reconstruction procedure for the effective action of
${\cal N}=4$ SYM theory can be realized completely for any terms in the expansion
(45), completing them by the corresponding terms containing
the hypermultiplet superfields.

\section{Summary}

We have presented the general approach to evaluation of the
one-loop effective action in ${\cal N}=2$ supersymmetric field
theories formulated in terms of ${\cal N}=1$ superfields. The
approach provides a calculation of the effective action by a
series in supercovariant derivatives with the coefficients
depending on the background superfields. The approach allows one
to reproduce the known results on the one-loop holomorphic and
non-holomorphic effective potentials depending on Abelian
background strengths in ${\cal N}=2$ SYM theories. We have studied
the structure of the low-energy effective action on non-Abelian
background superfields using a parametrically dependent family of
appropriate superfield $R_{\xi}$-gauges. For some values of the
gauge parameters, the non-Abelian non-holomor\-phic effective
potential is presented in an explicit form in terms of the the
Euler dilogarithm function. We have applied this general approach
to evaluation of the ${\cal N}=4$ supersymmetric low-energy
effective action in ${\cal N}=4$ SYM theory. We have found an
integral representation of the effective action for the constant
Abelian background strength, including the dependence on both the
${\cal N}=2$ gauge multiplet and the hypermultiplet superfields.
The four lowest terms of the
 effective action power expansion in the Abelian strength are
given in an explicit form.

\vspace{3mm}
{\large \hspace{-6mm}
\bf Acknowledgements }
\vspace{3mm}

The work was supported in part by INTAS grant, INTAS-00-00254 and
RFBR grant, project No 03-02-16193. I.L.Buchbinder is also
grateful to RFBR-DFG grant, project No 02-02-04002 and to DFG
grant, project No 436 RUS 113/669 and grant for Leading Russian
Scientific Schools, project No 1252.2003.2 for partial support.
The work of N.G.Pletnev and A.T.Banin was supported in part by
RFBR grant, project No 02-02-17884. A.T.B and N.G.P are very
grateful to the organizers of GRG11 for warm hospitality in Tomsk.

\end{document}